\documentclass[pra]{revtex4}
\usepackage{amssymb}
\usepackage{amsmath}
\usepackage{graphicx}
\begin{document}
\title{Shock Waves in Falling Coupled Harmonic Oscillators}
\author{Hidetsugu Sakaguchi}
\affiliation{Department of Applied Science for Electronics and Materials,
Interdisciplinary Graduate School of Engineering Sciences, Kyushu
University, Kasuga, Fukuoka 816-8580, Japan}
\begin{abstract}
Shock waves propagate in falling coupled harmonic oscillators.  
The bottom  end of coupled  harmonic oscillators does not fall downwards 
 until a shock wave reaches the bottom end. The exact solution can be expressed by the Fourier series expansion, and an approximate solution can be expressed by the integral of the Airy function. The width of the shock wave increases slowly in accordance with a power law.
\end{abstract}
\maketitle
Shock waves are generated in compressive fluids. Typical shock waves appear in  air compressed by supersonic planes or meteorites. 
There is a jump in the fields of pressure, temperature, and fluid velocity.
The Rankine-Hugoniot relation is satisfied for the jump under normal shock~\cite{rf:1}. There have been numerous investigations of shock waves~\cite{rf:2}. It is considered that a shock wave is a typical nonlinear wave. 
The simplest model of a shock wave is the Burgers equation~\cite{rf:3}. 
Nonlinearity and dissipation are essential for shock waves. 

We consider a linear chain of coupled harmonic oscillators under gravity. It is a linear system and there is no dissipation. It is a typical system of particles considered in a basic course of mechanics. However, there is a nontrivial phenomenon similar to a shock wave in this simple system. A similar phenomenon was discussed in falling elastic bars, using a partial differential equation~\cite{rf:4}.  We will focus on the effect of the discreteness in this paper.  

The model equation is written as 
\begin{eqnarray}
m\frac{d^2x_N}{dt^2}&=&k(-x_N+x_{N-1}+a)-mg,\nonumber\\
m\frac{d^2x_i}{dt^2}&=&k(x_{i+1}-2x_i+x_{i-1})-mg,\nonumber\\
m\frac{d^2x_1}{dt^2}&=&k(x_2-x_1-a)-mg,
\end{eqnarray}
where $N$ is the total number of particles of mass $m$, $k$ is the spring constant, $a$ is the natural length of the spring, $x_i$ is the height of the $i$th particle, and $g$ denotes the acceleration of gravity. 
The heights of the bottom and top particles are expressed respectively as  $x_1$ and  $x_N$. If $x_N$ is fixed to a constant value $x_{N0}$ by holding the top particle, the stationary positions of the other particles are determined from the relation $x_{i+1}-x_{i}=a+(mgi/k)$ as 
\begin{equation}
x_i=(i-1)a+\frac{mgi(i-1)}{2k}+x_1,
\end{equation}
where the position of the bottom particle, $x_1$, is expressed as $x_1=x_{N0}-(N-1)a-mgN(N-1)/(2k)$. We study the free-fall motion of this system by releasing the top particle with an initial velocity of 0 from the stationary state. 
There are five parameters, i.e., $m,k,a,g$, and $N$, in this system, but $N$ is the only essential parameter. The other parameters can be set to a unit value by changing the scales of $x$ and $t$.  

\begin{figure}[tbp]
\begin{center}
\includegraphics[height=4.cm]{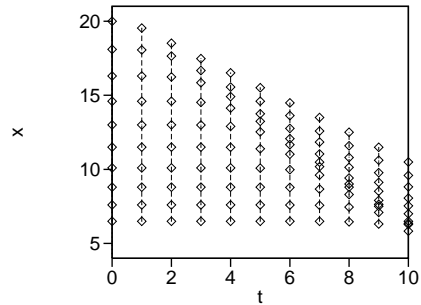}
\end{center}
\caption{Free fall of coupled harmonic oscillators of ten particles at $t=0,1,\cdots, 10$.  
}
\label{fig1}
\end{figure}
Figure 1 shows the positions of ten particles at $t=0,1,\cdots,10$ for $N=10,\, k=1,\,g=0.1,\,m=1$, and $a=1$.
The top particle falls with a nearly constant velocity. The bottom particle does  not move until $t\sim 8$ in the gravity field. This type of behavior is also observed in a falling slinky. The slinky is a toy of spring, that can walk downstairs. The interesting behavior of a falling slinky was studied by several authors using the wave equation for an elastic wave, which is a partial differential equation~\cite{rf:4,rf:5,rf:6}. 
The bottom particle does not move until the wave of deformation reaches the bottom, because information on the imbalance of force propagates with a finite velocity for an elastic wave.  

\begin{figure}[tbp]
\begin{center}
\includegraphics[height=4.cm]{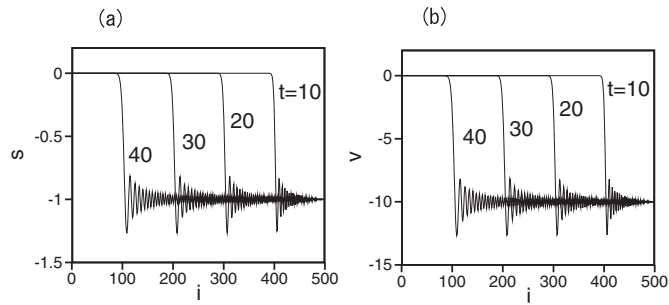}
\end{center}
\caption{Snapshot profiles of (a) elongation $s_i$ of spring and (b) velocity $v_i$ at $t=10,20,30$, and $40$.  
}
\label{fig2}
\end{figure}
Figure 2 shows four snapshot profiles of the elongation of the spring from the equilibrium state: $s_i=x_{i+1}-x_i-a-(mgi/k)$ and the velocity $v_i=dx_i/dt$ at $t=10,20,30$, and 40 for $k=100,a=1,m=1,g=0.2$, and $N=500$. Jumps appear in the profiles of $s_i$ and $v_i$, which are similar to a shock wave.
A shock wave propagates in the $-i$ direction with velocity $v_s=-10$.  This velocity is equal to the velocity of an elastic wave: $v_s=\sqrt{k/m}$. The particles in front of the shock wave are in the equilibrium state, and $s_i=0$ and $v_i=0$. The velocity of the particles behind the shock wave is nearly $v_d=-10$. The spring is compressed at the shock wave, and a jump of $1$ appears in the profile of $s_i$. The shock wave has a finite width, and damping oscillation is observed behind it.  It was shown analytically that a discontinuity similar to that of the shock wave propagates with the sound velocity using the wave equation previously~\cite{rf:4,rf:5,rf:6}, however, the discontinuity is unphysical. It is characteristic of our discrete system of coupled harmonic oscillators that a shock wave has a finite width and a tail structure of damping oscillation.

The total momentum $P=\sum_{i=1}^Nmv_i$ obeys the equation for the system of particles 
\begin{equation}
\frac{dP}{dt}=-mNg.
\end{equation}
If the particles behind a shock waves are assumed to have a constant velocity $v_d$ and there are $N_d$ particles in the region, the total momentum of this system is evaluated as $P=N_dmv_d$. Then,
\begin{equation}   
\frac{dN_d}{dt}mv_d=-mNg
\end{equation}
is satisfied. Because the shock wave propagates with velocity $v_s=\sqrt{k/m}$ and  the number of particles behind it increases with $dN_d/dt=v_s=\sqrt{k/m}$, the velocity $v_d$ is evaluated as
\begin{equation}
v_d=-\frac{Ng}{\sqrt{k/m}}.
\end{equation}
The evaluated velocity $v_d$ is -10 for $k=100,\,a=1,\,m=1,\,g=0.2$, and $N=500$, which is consistent with the numerical result. Equation (1) is rewritten as 
\begin{equation}
m\frac{dv_i}{dt}=k(s_{i}-s_{i-1}), \; \frac{ds_i}{dt}=v_{i+1}-v_{i}.
\end{equation}
Since $v_i\sim v_d$ far behind the shock wave, $s_{i}\sim s_{i-1}$ is satisfied. A jump appears in the profile of $s_i$ at the shock wave owing to the jump of $v_i$. The jump size of $s_i$ is evaluated to be at $v_d/v_s=-Ngm/k\sim -1$  using the relation $d s_i/dt\sim  (s_{i+1}-s_{i})/(1/v_s)=v_{i+1}-v_{i}$.  The whole profiles of $s_i$ and $v_i$ satisfy $s_i\sim (1/v_s)v_i$ from the same relation as that shown in Fig.~2. Here, $1/v_s$ is the time it takes for the shock wave to propagate by one particle. 
The relations of the jumps of the elongation of spring and the velocity correspond to the Rankine-Hugoniot relation for a shock wave in compressive fluids. 

\begin{figure}[tbp]
\begin{center}
\includegraphics[height=4.cm]{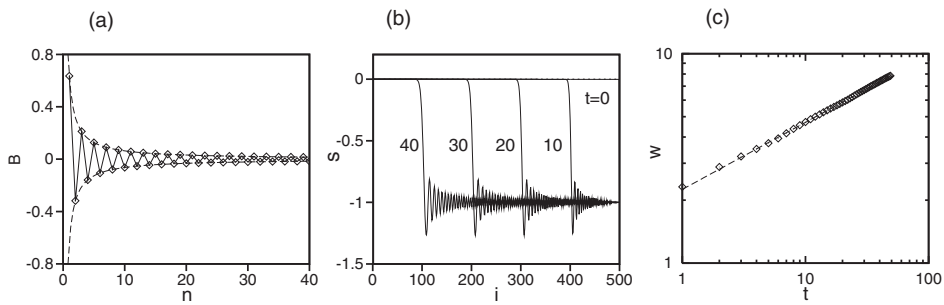}
\end{center}
\caption{(a) Fourier coefficients $B_n$ vs $n$. (b) Snapshot profiles of $s_i$ at $t=0,10,\cdots, 40$ obtained using eq.~(8). (c) Time evolution of the shock width $w(t)$ in a double-logarithmic plot. The dashed line is $w\sim t^{0.32}$.  
}
\label{fig3}
\end{figure}
 
The width of a shock wave cannot be evaluated using such a physical argument. However, the linear equation eq.~(1) can be exactly solved by the Fourier series expansion. The deviation $s_i$ from the stationary solution satisfies
\begin{equation}
m\frac{d^2s_i}{dt^2}=k(s_{i+1}-2s_i+s_{i-1}).
\end{equation}
Because of the definition $s_i=x_{i+1}-x_i-a-(mgi/k)$ and the boundary conditions for eq.~(1), expressed  as  $x_0=x_1-a$ and $x_{N+1}=x_N+a$, the boundary conditions for $s_i$ in eq.~(7) are expressed as $s_0=0$ and $s_{N}=-Nmg/k$. 
Note that the difference $|s_0-s_N|$ is equal to the jump size of $s_i$ at the shock wave shown in Fig.~2(a). 
The initial conditions for $s_i(t)$ are $ds_i/dt=0$,  $
s_i=0$ for $1\le i\le N-1$, and $s_i=-Nmg/k$ for $i=N$. Taking the boundary conditions into consideration, $s_i$ can be expanded using the Fourier series as
\begin{equation}
s_i(t)=-\frac{mgi}{k}+\sum_{n=1}^{N}\sin\frac{\pi n i}{N}\left (A_n\sin \omega_n t+B_n\cos\omega_n t\right ),
\end{equation}
where $A_n$ and $B_n$ are the Fourier coefficients. The frequency $\omega_n$ is given by 
\begin{equation}
\omega_n=\sqrt{\frac{2k}{m}\left (1-\cos\frac{\pi n}{N}\right )}=2\sqrt{\frac{k}{m}}\left |\sin\frac{\pi n}{2N}\right |.
\end{equation}
The Fourier coefficients $A_n$ are all zero from the initial condition $ds_i/dt=0$. The Fourier coefficients $B_n$ are calculated as
\begin{equation}
B_n=\frac{2}{N+1}\sum_{i=0}^N\sin\frac{\pi n i}{N}\left (s_i(0)+\frac{mgi}{k}\right ).
\end{equation}
Figure 3(a) shows the relationship of $B_n$ vs $n$. The dashed curves are $\pm 0.635/n$.
The Fourier coefficients $B_n\sim 0.635/n$ for odd $n$ and $B_n\sim -0.635/n$ for even $n$, when $n$ is relatively small. This is because $B_n$ are approximately evaluated for $n/N<<1$ as 
\[B_n\sim \frac{2}{N+1}\int_{0}^N\sin\frac{\pi n x}{N}\left (\frac{mgx}{k}\right )dx\sim (-1)^{n-1}\frac{2mgN}{\pi k n}\sim (-1)^{n-1}\frac{0.637}{n}.\]
Figure 3(b) shows $s_i(t)$ at $t=0,10,20,30$, and 40 obtained using eq.~(8). The numerical result shown in Fig.~2(a) is reproduced. We have estimated the width $w$ of the shock wave at the distance $i^{\prime}-i$ between two points satisfying $s_i(t)=-0.1$ and $s_{i^{\prime}}=-0.8$. $w$ is approximately calculated using the interpolation method, because our system is discrete and the position of $s_i=-0.1$ or -0.8 cannot be obtained exactly. Figure 3(c) shows the time evolution of $w(t)$ in a double-logarithmic plot. The width increases in accordance with a power law $w(t)\sim t^{\alpha}$, where $\alpha\sim 0.32$.  It increases owing to the dispersion of waves expressed by eq.~(9). If $\omega_n=\sqrt{k/m}(\pi n/N)$ is substituted into eq.~(8), the shock wave width is always 1 and the tail of the damping oscillation does not appear, or the shock wave is completely discontinuous. 

\begin{figure}[tbp]
\begin{center}
\includegraphics[height=4.cm]{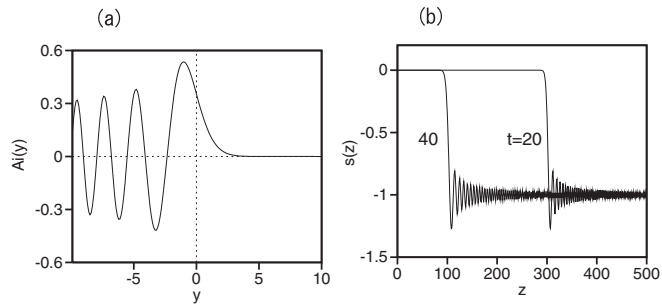}
\end{center}
\caption{(a) Airy function ${\rm Ai}(y)$, and (b) shock wave solution at $t=20$ and 40 obtained by eq.~(14).  
}
\label{fig4}
\end{figure}
There are two traveling waves in a system described by eq.~(7), i.e., downward and upward waves. If only downward waves are taken into consideration, and the dispersion relation eq.~(9) is approximated as $\omega(q)=2\sqrt{k/m}|\sin(q/2)|\sim cq-\beta q^3$ with $q=\pi n/N$, $c=\sqrt{k/m}$, and $\beta=c/24$, the linearized Kortweg-de Vries (KdV) equation
\begin{equation}
\frac{\partial s}{\partial t}=c\frac{\partial s}{\partial z}+\beta\frac{\partial^3s}{\partial z^3}
\end{equation}
is derived by the continuum approximation of eq.~(7).
The general solution of this equation can be expressed via the Airy function Ai$(x)$ as ~\cite{rf:7}
\begin{equation}
s(z,t)=(3\beta t)^{-1/3}\int_{-\infty}^{\infty}{\rm Ai}\left \{\frac{-z-ct+z^{\prime}}{(3\beta t)^{1/3}}\right \}s(z^{\prime},0)dz^{\prime},
\end{equation} 
where $s(z,0)$ is the initial value of $s(z,t)$ at $t=0$.
The Airy function can be expressed with the integral form
\begin{equation}
{\rm Ai}(y)=\frac{1}{\pi}\int_0^{\infty}\cos\left (\frac{x^3}{3}+xy\right )dx.
\end{equation}
Figure 4(a) shows the Airy function ${\rm Ai}(y)$. Owing to the initial condition, $u(z,0)=-Nmg/k$ for $z\ge N$ and $u(z,0)=0$ for $z<N$. The substitution of this initial condition into eq.~(12) yields
\begin{equation}
s(z,t)=(3\beta t)^{-1/3}\frac{(-Nmg)}{k}\int_{N}^{\infty}{\rm Ai}\left \{\frac{-z-ct+z^{\prime}}{(3\beta t)^{1/3}}\right \}dz^{\prime}=\frac{(-Nmg)}{k}\int_{(N-z-ct)/(3\beta t)^{1/3}}^{\infty}{\rm Ai}(y)dy.
\end{equation}
That is, the solution $s(z,t)$ can be expressed with the integral of the Airy function and has a scaling form of $s(z,t)=\tilde{s}((N-z-ct)/(3\beta t)^{1/3})$.  Figure 4(b) shows the profiles of $s(z)$ at $t=20$ and $t=40$ calculated using eq.~(14) with $c=10$ and $\beta=c/24=5/12$. The results in Figs.~2(a) and 3(b) are reproduced very well.  A shock wave propagates with velocity $c$, and  the width of the shock wave increases as $t^{\alpha}$ with $\alpha=1/3$ owing to the dispersion effect. The tail structure of damping oscillation is due to the form of the Airy function Ai$(y)$ in the region of $y<0$.  

To summarize, we have shown a phenomenon similar to a shock wave in falling coupled harmonic oscillators. The jumps in the profiles of the velocity and the elongation of spring are evaluated. The shock wave solution can be exactly solved by the Fourier series expansion. The shock wave has a finite width and a tail of damping oscillation, and the width increases with time. This is  different  from the case of a continuous system of a falling elastic bar previously studied. The shock wave solution can be further approximated by the integral of the Airy function,  which is a solution of the linearized KdV equation.  

The solution can be mathematically solved; however, it is physically counterintuitive that particles behind a shock wave have a nearly constant velocity in the gravity field similar to the terminal velocity determined by the viscosity. The harmonic oscillators behind the shock wave are in an equilibrium state of forces, even though the oscillators are accelerated by gravity and compressed by the shock wave. The tail of damping oscillation is considered to be the harmonic oscillation of particles induced by compression by the shock wave, however, the mechanism of the damping remains to be clarified. 

Our model system is very simple but shows unexpected behavior. Our model might be an instructive model in a basic course of mechanics. We expect that our simple model to be applied to the study of phenomena such as avalanche snowslides or landslides falling along a slope by incorporating the effect of friction or some other effects.


\begin{thebibliography}{99}
\bibitem{rf:1} W.~J.~M.~Rankine: Philos. Trans. Roy. Soc. London {\bf 160} (1870) 277.
\bibitem{rf:2} e.g., I.~I.~Glass: {\it Shock Wave and Man},  (Toronto Univ. Press, Toronto,1974).
\bibitem{rf:3} J.~M.~Burgers: {\it The Nonlinear Diffusion Equation} (D. Reidel Publishing Company, Dordrecht-Boston, 1974).
\bibitem{rf:4} J.~M.~Aguirregabiria, A.~Hern\'andez, and M.~Rivas: Am. J. Phys. {\bf 75} (2007) 583.
\bibitem{rf:5} M.~G.~Calkin: Am. J. Phys. {\bf 61} (1993) 261.
\bibitem{rf:6} R.~C.~Cross and M.~S.~Wheatland: Am. J. Phys. {\bf  80} (2012) 1051.
\bibitem{rf:7} e.g., V.~I.~Karpman: {\it Nonlinear Waves in Dispersive Media} (Pergamon, Oxford, New York, 1975). 

\end{thebibliography}
\end{document}